\documentclass[aps,prc,preprint,amsmath,amssymb,showpacs,preprintnumbers,superscriptaddress]{revtex4-1}
\usepackage{CJK}
\usepackage{graphicx}
\usepackage{dcolumn}
\usepackage{bm}
\usepackage{color}
\usepackage{hyperref}
\allowdisplaybreaks[4]

\begin{document}

\title{Configuration interaction projected density functional theory: effects of four-quasiparticle configurations and time-odd interactions}
\author{Y. K. Wang}
\affiliation{State Key Laboratory of Nuclear Physics and Technology, School of Physics, Peking University, Beijing 100871, China}

\author{P. W. Zhao}
\email{pwzhao@pku.edu.cn}
\affiliation{State Key Laboratory of Nuclear Physics and Technology, School of Physics, Peking University, Beijing 100871, China}

\author{J. Meng}
\email{mengj@pku.edu.cn}
\affiliation{State Key Laboratory of Nuclear Physics and Technology, School of Physics, Peking University, Beijing 100871, China}

\begin{abstract}
  The effects of four-quasiparticle configurations and time-odd interactions are investigated in the framework of configuration interaction projected density functional theory by taking the yrast states of $^{60}$Fe as examples.
  Based on the universal PC-PK1 density functional, the energies of the yrast states with spin up to $20\hbar$ and the available $B(E2)$ transition probabilities are well reproduced.
  The yrast states are predicted to be of four-quasiparticle structure above spin $I = 16\hbar$.
  The inclusion of the time-odd interactions increases the kinetic moments of inertia and delays the appearance of the first band crossing, and, thus, improves the description of the data.
\end{abstract}

\maketitle

\section{Introduction}
The configuration interaction shell model (CI-SM)~\cite{Caurier2005Rev.Mod.Phys.427488} and the nuclear density functional theory (DFT)~\cite{Bender2003Rev.Mod.Phys.75121--180,Meng2016,Meng2021AAPPSBulletin17} are widely used approaches for the description of nuclear properties.
The CI-SM describes nuclear spectroscopic properties in the laboratory frame and captures both the excitation of single particles and the collective correlations coming from the coherent motion of many nucleons.
Because of the limitation of its model space, it is usually applied to light and medium heavy nuclei.
The nuclear DFT is based on the idea that the ground-state energy of a nucleus can be expressed as a functional of the nucleon density, and is applicable for nuclei all over the nuclide chart.
The nuclear DFT takes into account many-body correlations by breaking essential symmetries and describes fruitful physics around the minima of the potential energy surface.
However, the nuclear DFT, in the first place, is limited to describe nuclear ground states.
For quantitative investigations of nuclear spectroscopic properties, one needs to resort to proper extensions based on the DFT.

During the past decades, great efforts have been made to extend the nuclear DFT for nuclear spectroscopic properties.
The generator coordinate method (GCM)~\cite{Ring2004} which includes the quantum fluctuations along some relevant collective variables, is one of the effective approaches.
The GCM approach has been implemented in nonrelativistic~\cite{Valor2000NuclearPhysicsA145164,Bender2006Phys.Rev.C034322,
RodriguezGuzman2002NuclearPhysicsA201235} and relativistic~\cite{Niksic2006Phys.Rev.C034308,Nikifmmodecheckselsevsfiiifmmodeacutecelsecfi2006Phys.Rev.C064309} DFTs,
and has achieved great successes in the shape coexistence~\cite{Rodriguez2011PhysicsLettersB255259,Yao2014Phys.Rev.C054306,Egido2020Phys.Rev.Lett.192504}, the erosion of shell-closure~\cite{Rodriguez2011Phys.Rev.C051307}, the collective band structures of superheavy nuclei~\cite{Egido2021Phys.Rev.Lett.192501}, the novel modes of nuclear weak decays~\cite{Rodriguez2010Phys.Rev.Lett.105252503,Song2014Phys.Rev.C90054309}, etc.
Nevertheless, the DFT-based GCM approaches are limited to describe the spectra with energies below the first two-quasiparticle excitations~\cite{Niksic2011ProgressinParticleandNuclearPhysics519548}.
For the description of spectra with higher excitation energies, the cranking model~\cite{Inglis1956Phys.Rev.17861795} based on nonrelativistic~\cite{Egido1993Phys.Rev.Lett.28762879,Satula2004ReportsonProgressinPhysics131200} and relativistic~\cite{Vretenar2005PhysicsReports409101-259,Meng2013Front.Phys.855--79} DFTs
have been demonstrated as powerful tools.
Moreover, the two-dimensional and the three-dimensional tilted axis cranking DFTs have been successfully applied to many novel rotational phenomena~\cite{Olbratowski2004Phys.Rev.Lett.93052501,Peng2008Phys.Rev.C78024313,
Zhao2011Phys.Lett.B699181--186,Zhao2011Phys.Rev.Lett.107122501,Zhao2017PhysicsLettersB7731-5}.
However, the band crossing phenomena cannot be properly described by the cranking approach~\cite{Hamamoto1976NuclearPhysicsA1528} because the cranking states are obtained at the constant rotational frequency rather than the constant angular momentum.
Recent progresses on considering the correlations due to the symmetry breaking and mixing states beyond the rotating mean field have been made based on the nonrelativistic DFTs~\cite{Egido2016Phys.Rev.Lett.116052502,Shimada2016Phys.Rev.C044317}.

To describe the nuclear spectroscopic properties with beyond mean-field correlations up to high spin, the configuration interaction projected density functional theory (CI-PDFT) was proposed in Ref.~\cite{Zhao2016Phys.Rev.C94041301}.
The starting point of CI-PDFT is a well-defined density functional whose self-consistent solution corresponds to a state at the energy minimum of the potential energy surface and contains already important physics.
The additional correlations required for the description of spectroscopic properties are considered with shell-model calculations in a configuration space built on top of the self-consistent solution.
Due to the fact that such a configuration space is constructed with the optimized bases, its dimension is much smaller than that in the traditional shell models.
The CI-PDFT could be used to describe the nuclear spectroscopic properties up to high spin without additional parameters beyond the well-defined density functional.
The CI-PDFT has been later implemented in the framework of nonrelativistic DFTs~\cite{Satula2016Phys.Rev.C024306,Konieczka2018Phys.Rev.C034310} where the authors use an alternative name, i.e., no-core configuration-interaction approach rooted in multireference DFT.

In the previous application of the CI-PDFT~\cite{Zhao2016Phys.Rev.C94041301}, the four-quasiparticle configurations are not included.
Moreover, the time-odd parts of the density functional are neglected for simplicity.
As stated in Ref.~\cite{Zhao2016Phys.Rev.C94041301}, the consideration of time-odd components and four-quasiparticle configurations would give better agreements with the experimental data for higher spin states.
In fact, the investigations within the projected shell model (PSM) based on the phenomenological pairing-plus-quadrupole interaction, which includes the four-quasiparticle configurations, can reproduce the higher spin states well~\cite{Sun2009Phys.Rev.C054306}.
In the present work, the CI-PDFT including contributions from the four-quasiparticle configurations and time-odd interactions is developed.
By taking the yrast states of $^{60}$Fe populated up to spin $I = 20\hbar$~\cite{Deacon2007Phys.Rev.C054303} as examples, the effects of four-quasiparticle configurations and time-odd interactions are discussed.

\newpage

\section{Theoretical framework}~\label{sec.II}
The starting point of the point-coupling relativistic density functional is the following Lagrangian density~\cite{Meng2016},
\begin{equation}\label{eq:lagrangian}
  \begin{split}
    \mathcal{L} &= \bar{\psi}(i\gamma_\mu\partial^\mu - m)\psi\\
    &-\frac{1}{2}\alpha_S(\bar{\psi}\psi)(\bar{\psi}\psi) - \frac{1}{2}\alpha_V(\bar{\psi}\gamma_\mu\psi)(\bar{\psi}\gamma^\mu\psi) -\frac{1}{2}\alpha_{TV}(\bar{\psi}\vec{\tau}\gamma_\mu\psi)(\bar{\psi}\vec{\tau}\gamma^\mu\psi)\\
    & -\frac{1}{3}\beta_S(\bar{\psi}\psi)^3 - \frac{1}{4}\gamma_S(\bar{\psi}\psi)^4 - \frac{1}{4}\gamma_V[(\bar{\psi}\gamma_\mu\psi)(\bar{\psi}\gamma^\mu\psi)]^2\\
    &-\frac{1}{2}\delta_S\partial_\nu(\bar{\psi}\psi)\partial^\nu(\bar{\psi}\psi) - \frac{1}{2}\delta_V\partial_\nu(\bar{\psi}\gamma_\mu\psi)\partial^\nu(\bar{\psi}\gamma^\mu\psi)
    -\frac{1}{2}\delta_{TV}\partial_\nu(\bar{\psi}\vec{\tau}\gamma_\mu\psi)\partial^\nu(\bar{\psi}\vec{\tau}\gamma^\mu\psi)\\
    &-\frac{1}{4}F^{\mu\nu}F_{\mu\nu} - e\frac{1-\tau_3}{2}\bar{\psi}\gamma^\mu\psi A_\mu,
  \end{split}
\end{equation}
which contains 9 coupling constants $\alpha_S$, $\alpha_V$, $\alpha_{TV}$, $\beta_S$, $\gamma_S$, $\gamma_V$, $\delta_S$, $\delta_V$, and $\delta_{TV}$.
The subscripts $S$, $V$, and $TV$ represent scalar, vector, and isovector, respectively.
It is clear that the terms associated with the coupling constants are written as a power series in $\bar{\psi}\mathcal{O}\Gamma\psi$ and their derivatives, where $\mathcal{O}\in\{1,\vec{\tau}\}$ and $\Gamma\in\{1,\gamma_\mu\}$, and Greek indices $\mu$ run over the Minkowski indices $0$, $1$, $2$, and $3$.
Note that the time-odd interactions originate from the space-like components, i.e., the components with $\mu = 1$, $2$, and $3$, of the vector and isovector channels.
From the Lagrangian density, one can derive the Hamiltonian by Legendre transformation.
Details can be seen in Ref.~\cite{Meng2016}.

For open-shell nuclei, a unified treatment of both the mean field and pairing correlations is realized by solving the following relativistic Hartree-Bogoliubov (RHB) equation~\cite{Meng2016},
\begin{equation}\label{eq:RHB}
  \left(\begin{array}{cc}
    \hat{h}_D-\lambda& \hat{\Delta}\\
    -\hat{\Delta}^\ast & -\hat{h}_D^\ast + \lambda
  \end{array}\right)
  \left(\begin{array}{c}
    U_k\\
    V_k
  \end{array}\right) = E_k
  \left(\begin{array}{c}
    U_k\\
    V_k
  \end{array}\right).
\end{equation}
Here, $U_k$ and $V_k$ are the quasiparticle (qp) wavefunctions, $\hat{h}_D$ is the single-particle Dirac Hamiltonian
\begin{equation}\label{eq:Dirac}
  \hat{h}_D = \bm{\alpha}\cdotp(\bm{p}-\bm{V}) + \beta(m + S) + V,
\end{equation}
and $\lambda$ is the Fermi surface.
The scalar field $S$ and vector field $V^\mu$ are connected in a self-consistent way to various densities and currents~\cite{Meng2016}.
The matrix element of the pairing field $\hat{\Delta}$ is
\begin{equation}\label{eq:pair-matrix}
  \Delta_{ab} = \frac{1}{2}\sum_{c,d}\langle ab|V^{pp}|cd\rangle_a\kappa_{cd},
\end{equation}
where $V^{pp}$ is the pairing force, and $\kappa$ is the pairing tensor determined by the qp wavefunctions.
In the present work, $V^{pp}$ is chosen as the finite-range separable force proposed in Ref.~\cite{Tian2009Phys.Lett.B67644--50}.
By solving the RHB equation \eqref{eq:RHB} iteratively, one can obtain the intrinsic ground state $|\Phi_0\rangle$, together with a set of qp orbits $i$ defined by
\begin{equation}
  \hat{\beta}_i|\Phi_0\rangle = 0,\quad \mathrm{for}~\forall~i,
\end{equation}
in which $\hat{\beta}_i$ represents the qp annihilation operator~\cite{Ring2004}.

Starting from the intrinsic ground state $|\Phi_0\rangle$, one can construct the multi-qp configurations.
For example, the multi-qp states up to four-qp configurations for even-even nuclei are
\begin{equation}\label{eq:config-space}
  \{|\Phi_0\rangle, \hat{\beta}^\dag_{\nu_i}\hat{\beta}^\dag_{\nu_j}|\Phi_0\rangle, \hat{\beta}^\dag_{\pi_i}\hat{\beta}^\dag_{\pi_j}|\Phi_0\rangle, \hat{\beta}^\dag_{\nu_i}\hat{\beta}^\dag_{\nu_j}\hat{\beta}^\dag_{\pi_k}\hat{\beta}^\dag_{\pi_l}|\Phi_0\rangle,
  \hat{\beta}^\dag_{\pi_i}\hat{\beta}^\dag_{\pi_j}\hat{\beta}^\dag_{\pi_k}\hat{\beta}^\dag_{\pi_l}|\Phi_0\rangle,
  \hat{\beta}^\dag_{\nu_i}\hat{\beta}^\dag_{\nu_j}\hat{\beta}^\dag_{\nu_k}\hat{\beta}^\dag_{\nu_l}|\Phi_0\rangle\}.
\end{equation}
These states form the configuration space of the CI-PDFT.
In general, multi-qp configurations with half protons and half neutrons are energy favored.
Therefore, four-qp configurations with all like particles could be excluded if one is interested in only the yrast states~\cite{HARA1995InternationalJournalofModernPhysicsE04637-785,Wang2016Phys.Rev.C93034322}.

The wavefunction $|\Psi^\sigma_{IM}\rangle$ in the laboratory frame with good angular momentum quantum numbers $I$ and $M$ is written as~\cite{Zhao2016Phys.Rev.C94041301}
\begin{equation}
  |\Psi^\sigma_{IM}\rangle = \sum_\eta F^{I\sigma}_\eta\hat{P}^I_{MK}|\Phi_\eta\rangle.
\end{equation}
Here, $\eta$ runs over the intrinsic states taken from the configuration space in Eq.~\eqref{eq:config-space}, and $\hat{P}^I_{MK}$ is the angular momentum projection operator.
Similar to Ref.~\cite{Zhao2016Phys.Rev.C94041301}, each multi-qp state $|\Phi_\eta\rangle$ is assumed to be axially deformed and is characterized by the total intrinsic magnetic quantum number $K$.
The expansion coefficients $F^{I\sigma}_{\eta}$ are determined by the diagonalization of the Hamiltonian in a shell-model subspace spanned by the nonorthogonal projected bases $\{\hat{P}^I_{MK}|\Phi_\eta\rangle\}$, or equivalently by solving the following generalized eigenvalue equation
\begin{equation}\label{eq:Hill-wheeler}
  \sum_{\eta'}[H^I_{\eta\eta'} - E^{I\sigma}N^I_{\eta\eta'}]F^{I\sigma}_{\eta'} = 0.
\end{equation}
The energy kernel $H^I_{\eta\eta'}$ and the norm matrix $N^I_{\eta\eta'}$ are defined respectively as
\begin{equation}
  H^I_{\eta\eta'} = \langle\Phi_\eta|\hat{H}\hat{P}^I_{KK'}|\Phi_{\eta'}\rangle,\quad N^I_{\eta\eta'} = \langle\Phi_\eta|\hat{P}^I_{KK'}|\Phi_{\eta'}\rangle.
\end{equation}
Here, $\hat{H}$ represents the Hamiltonian derived from the Lagrangian density~\eqref{eq:lagrangian} including the pairing term.
The energy kernel and norm matrix are calculated by the Pfaffian algorithms proposed in Refs.~\cite{Bertsch2012Phys.Rev.Lett.108042505,Hu2014PhysicsLettersB734162-166}.
In the present work, the particle number projection is not included, and an approximate scheme to correct the mean value of the particle number is used~\cite{Yao2011Phys.Rev.C014308}.
The time-odd interactions associated with the space-like components of vector and isovector channels are considered.

The generalized eigenvalue equation \eqref{eq:Hill-wheeler} is solved by the standard procedure.
Once the expansion coefficients $F^{I\sigma}_{\eta}$ are known, one can construct the collective wavefunctions~\cite{Ring2004}
\begin{equation}\label{Eq:G-wave}
  G^{I\sigma}_\eta = \sum_{\eta'}(N^I)^{1/2}_{\eta\eta'}F^{I\sigma}_{\eta'},
\end{equation}
which are interpreted as the probability amplitudes of the states $|\Phi_\eta\rangle$.
The reduced $E2$ transition probabilities between an initial state $I_i$ and a final state $I_f$ is defined by
\begin{equation}
  B(E2: I_i,\sigma\rightarrow I_f,\sigma') = \frac{e^2}{2I_i+1}|\langle\Psi^{\sigma'}_{I_f}||\hat{Q}_2||\Psi^{\sigma}_{I_i}\rangle|^2.
\end{equation}
Note that $e$ denotes the bare value of the proton charge, and there is no need to introduce the effective charges in the CI-PDFT calculations~\cite{Zhao2016Phys.Rev.C94041301}.

\section{Numerical details}~\label{sec.III}
In the present work, the CI-PDFT with time-odd interactions and four-qp configurations are illustrated by taking the yrast states of $^{60}$Fe~\cite{Deacon2007Phys.Rev.C054303} as examples.
The relativistic point-coupling Lagrangian PC-PK1~\cite{Zhao2010Phys.Rev.C82054319} is adopted for both the mean-field and CI-PDFT calculations.
For the pairing channel, the finite-range separable force~\cite{Tian2009Phys.Lett.B67644--50} with pairing strength $G = 728$ $\mathrm{MeV}\cdotp\mathrm{fm}^3$ is used.
The RHB equation \eqref{eq:RHB} is solved self-consistently by expanding the qp wavefunctions in terms of three-dimensional harmonic oscillator basis in Cartesian coordinates~\cite{Niksic2014Comput.Phys.Commun.1851808} with $N_f = 10$ major shells.
The configuration space for the present CI-PDFT calculations consists of zero-, two- and four-qp states.
For simplicity, the four-qp states are constructed by coupling two-qp proton states with two-qp neutron states, and those with four like-nucleon configurations are not included in the present investigation.

\section{Results and discussion}~\label{sec.IV}
\begin{figure}[h!]
  \centering
  \includegraphics[width=0.5\textwidth]{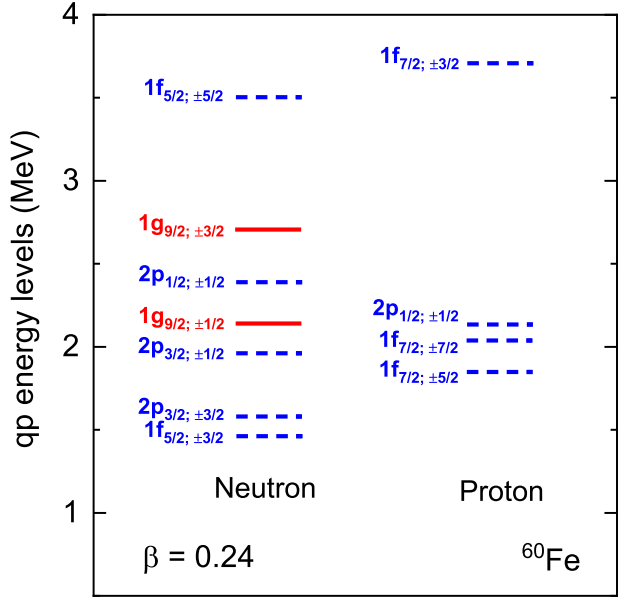}
  \caption{(Color online) The neutron and proton qp energy levels of $^{60}$Fe.
  The qp energy levels with positive and negative parities are denoted respectively by solid and dashed lines.
  The approximate spherical quantum numbers are used to label the qp energy levels (see text).}
  \label{Fig:qp-levels}
\end{figure}
In Fig.~\ref{Fig:qp-levels}, the neutron and proton qp energy levels with energy lower than 4.0 MeV are depicted.
The qp energy levels are obtained by solving the RHB equation~\eqref{eq:RHB} iteratively, and the obtained ground state is deformed with quadrupole deformation $\beta = 0.24$.
Therefore, the spherical quantum numbers $nljm$ used to label the qp levels in Fig.~\ref{Fig:qp-levels} are only the main spherical components of the deformed qp wavefunctions.
Due to the time-reversal symmetry, the qp energy levels are twofold degenerate, and the energies of qp levels denoted by quantum numbers with opposite projections on the third axis are the same.
For the mean-field calculation, the time-odd interactions do not contribute due to the time-reversal symmetry.
The corresponding creation operators $\hat{\beta}^\dag$ of the qp states are then used to generate the qp excited states~$|\Phi_\eta\rangle$.
As shown in Eq.~\eqref{eq:config-space}, the $|\Phi_\eta\rangle$ together with the intrinsic ground state $|\Phi_0\rangle$ form the configuration space of the CI-PDFT.

The dimension of the configuration space is truncated with a qp excitation energy cutoff $E_{\mathrm{cut}}$.
The adopted $E_{\mathrm{cut}}$ for two-qp states here is $5.0$ MeV, and the resultant configuration space consists of 30 states including 21 two-qp states for neutron and 9 ones for proton.
For the four-qp states, only the ones with two $g_{9/2}$ neutrons are considered, and the number of four-qp states in the configuration space is 27.

\begin{figure}[h!]
  \centering
  \includegraphics[width=0.5\textwidth]{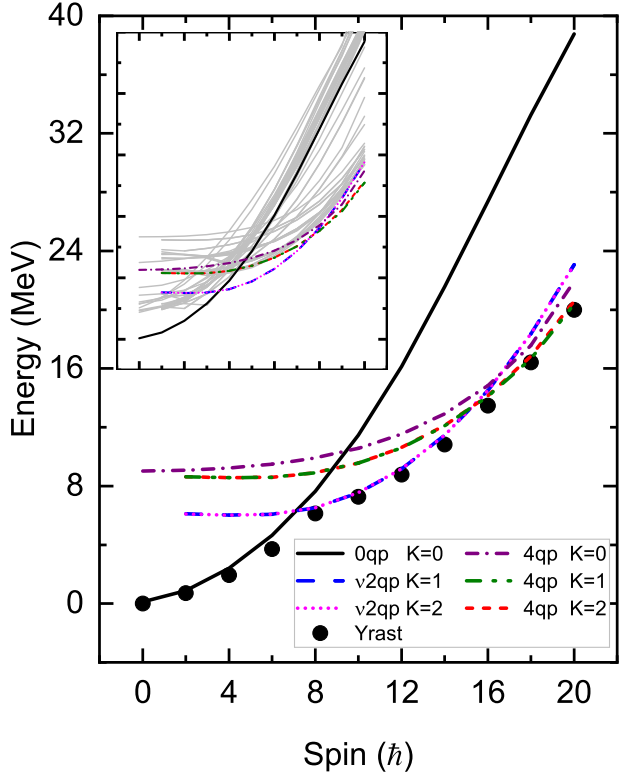}
  \caption{(Color online) The calculated yrast states (solid dots) and energies of angular momentum projected states for some important configurations (different lines).
  Inset: The energies of angular momentum projected states for all the configurations.
  Note that the energy of the yrast state $0^+$ is renormalized to zero.}
  \label{Fig:band-diagram}
\end{figure}

In the following, all the forementioned 58 states including 30 two-qp states, 27 four-qp states as well as one intrinsic ground state $|\Phi_0\rangle$ are included in the CI-PDFT calculations.
The energies of angular momentum projected states for the 58 configurations as functions of spin are plotted in the inset of Fig.~\ref{Fig:band-diagram}.
Such a plot is also named as the band diagram~\cite{HARA1995InternationalJournalofModernPhysicsE04637-785}, which provides
useful information for understanding the structure changes even before the diagonalization is carried out.
The energy of a band $\eta$ is defined as,
\begin{equation}
  E_\eta(I) = \frac{\langle\Phi_\eta|\hat{H}\hat{P}^I_{KK}|\Phi_\eta\rangle}{\langle\Phi_\eta|\hat{P}^I_{KK}|\Phi_\eta\rangle},
\end{equation}
which represents the expectation value of the Hamiltonian with respect to the projected qp states $|\Phi_\eta\rangle$.
Because the configurations are axially deformed in the present work, it is convenient to use the quantum number $K$ to classify them.

In Fig.~\ref{Fig:band-diagram}, the zero-qp band associated with the intrinsic ground state $|\Phi_0\rangle$ (0qp $K=0$) and the bands built on top of five additional two- or four-qp configurations that represent the most important ones for each kind of configuration are presented.
The two-qp neutron bands are marked as ``$\nu$2qp", and the corresponding configurations are respectively $\nu(g_{9/2,-1/2})^1(g_{9/2,3/2})^1$ for $K=1$ and $\nu(g_{9/2,1/2})^1(g_{9/2,3/2})^1$ for $K=2$.
The four-qp bands marked as ``4qp" are coupled with two $g_{9/2}$ neutrons and two $(fp)$ protons, and their configurations are $\nu(g_{9/2,-1/2})^1(g_{9/2,1/2})^1\otimes\pi(p_{1/2,1/2})^1(p_{1/2,-1/2})^1$ for $K=0$, $\nu(g_{9/2,-1/2})^1(g_{9/2,3/2})^1\otimes\pi(p_{1/2,1/2})^1(p_{1/2,-1/2})^1$ for $K=1$, and $\nu(g_{9/2,1/2})^1(g_{9/2,3/2})^1\otimes\pi(p_{1/2,1/2})^1(p_{1/2,-1/2})^1$ for $K=2$, respectively.
It is seen that the energy of the 0qp band with $K=0$ increases with spin and the band quickly enters into the high energy region, thus becoming unfavoured for the states in the high spin region.
In contrast, the $\nu2$qp bands with $K=1$ and $K=2$ show first a constant dependence with spin, and then cross with the 0qp band at $I=8\hbar$.
This behavior makes the two-qp states with two $g_{9/2}$ neutrons the most important configurations for the spin interval $I=8\sim14\hbar$.
The $\nu2$qp bands become less favorable above $I=14\hbar$, and the 4qp bands with $K=0$, $K=1$, and $K=2$ cross them and become more important.
From the analysis of the band diagram, the dominated role of the $g_{9/2}$ neutrons for the description of the yrast states of $^{60}$Fe is clearly illustrated.

\begin{figure}[h!]
  \centering
  \includegraphics[width=0.5\textwidth]{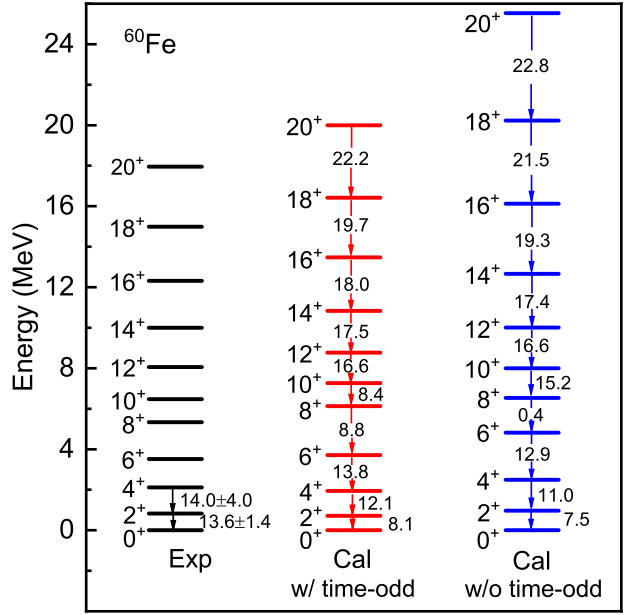}
  \caption{(Color online) The energy spectra (MeV) and $B(E2)$ transition probabilities (W.u.) calculated by the CI-PDFT for the yrast band in $^{60}$Fe, in comparison with data available in Refs.~\cite{Deacon2007Phys.Rev.C054303,Browne2013NuclearDataSheets18492022}.}
  \label{Fig:Energy-spectra}
\end{figure}
\begin{figure}[h!]
  \centering
  \includegraphics[width=0.7\textwidth]{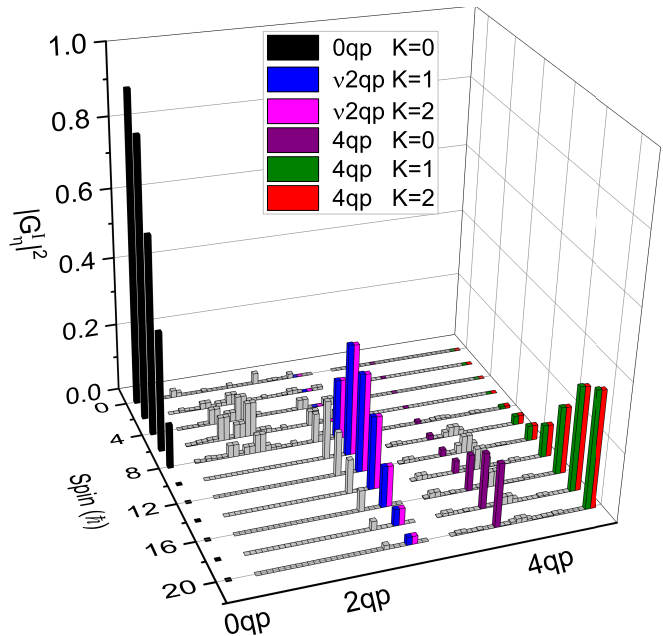}
  \caption{(Color online) The probability amplitudes $|G^I_\eta|^2$ for all the qp states involved in the description of the yrast states.
  For each spin, the $|G^I_\eta|^2$ are divided into three regions and are marked as ``0qp", ``2qp", and ``4qp", representing the contributions from intrinsic ground state, two-qp states, and four-qp states, respectively.
  The results of two-qp and four-qp states are plotted according to their qp excitation energies in ascending order.
  The $|G^I_\eta|^2$ associated with the important qp states are denoted by different color bars.}
  \label{Fig:collective-wave}
\end{figure}

The solid dots marked as ``Yrast" in Fig.~\ref{Fig:band-diagram} denote the lowest energy states for a given spin $I$, which are obtained by mixing all the angular momentum projected states through the diagonalization of the Hamiltonian.
These are the theoretical results to be compared with the data.
The energies and $B(E2)$ transition probabilities for the yrast states in $^{60}$Fe calculated by the CI-PDFT with and without time-odd interactions and their comparisons with the available data~\cite{Deacon2007Phys.Rev.C054303,Browne2013NuclearDataSheets18492022} are shown in Fig.~\ref{Fig:Energy-spectra}.
Meanwhile, we show the probability amplitudes $|G_\eta^I|^2$ for each of the yrast states with spin $I$ in Fig.~\ref{Fig:collective-wave} to better illustrate the band structure changes.

It is found that the energy levels as well as the $E2$ transition probabilities calculated by the CI-PDFT with time-odd interactions agree well with the available data.
In particular, the irregularity shown in the energy levels, i.e., the compressed levels spacings, at around $I=8\hbar$ is also reproduced satisfactorily.
The irregularity is mainly caused by the band crossing between the $\nu2$qp neutron bands with $K=1, 2$ and the 0qp band with $K=0$.
From the probability amplitudes $|G_\eta^I|^2$ shown in Fig.~\ref{Fig:collective-wave}, it is seen that the 0qp state plays dominated roles for the yrast states with $I\leq6\hbar$.
The corresponding probability amplitude $|G_\eta^I|^2$ is 0.90 for state with $I = 0\hbar$ and then decreases gradually with spin, being 0.13 for the yrast state with $I=8\hbar$.
The $|G^I_\eta|^2$ of the $\nu2$qp neutron states with $K=1$ and $2$ increase suddenly at $I=8\hbar$, which indicates that the yrast band structure changes from the 0qp state to the $\nu2$qp states and is consistent with the drop observed in the $B(E2)$ value at around $I=8\hbar$.
The $\nu2$qp neutron states hold their dominated roles up to $I = 14\hbar$, after which the 4qp states with $K=0$, $K=1$, and $K=2$ begin to compete with them.
The 4qp states win after $I=16\hbar$, and the yrast states are predicted to be of 4qp structure.
Note that the second band crossing caused by the 4qp states is much weaker than the one caused by the $\nu2$qp states, and, thus, no clear drops of $B(E2)$ values are observed at around $I=14\hbar$.

\begin{figure}[h!]
  \centering
  \includegraphics[width=0.5\textwidth]{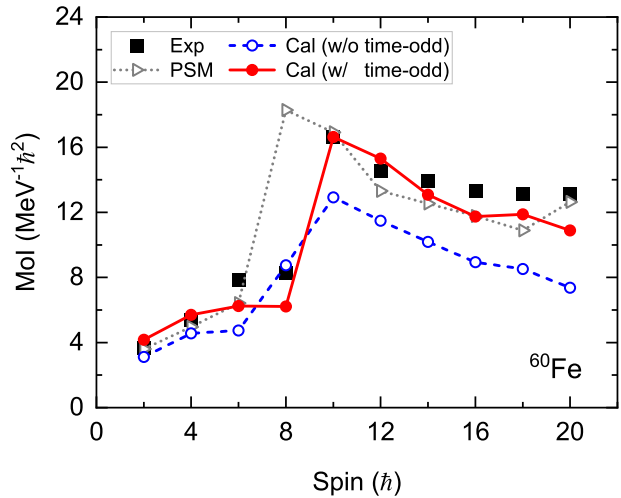}
  \caption{(Color online) The kinetic moment of inertia calculated by the CI-PDFT with and without time-odd interactions, in comparison with the data~\cite{Deacon2007Phys.Rev.C054303} and the results calculated by the PSM~\cite{Sun2009Phys.Rev.C054306}.}
  \label{Fig:moment of inertia}
\end{figure}

The energies of yrast states are overestimated by the CI-PDFT calculations without time-odd interactions.
The excitation energy for the yrast state with $I=20\hbar$ is predicted to be 25.5 MeV, around 7.5 MeV higher than the corresponding data.
The discrepancy might originate from the underestimation of the corresponding moments of inertia.
To better clarify this phenomenon, the kinetic moments of inertia (MOI) calculated by the CI-PDFT with and without time-odd interactions are shown in Fig.~\ref{Fig:moment of inertia}, in comparison with the data and the results calculated by the PSM.

It is seen that the MOI calculated without time-odd interactions underestimate the data, especially for the ones with $I\geq10\hbar$.
In comparison, the PSM results in Ref.~\cite{Sun2009Phys.Rev.C054306} reproduce the data well for states with $I=0\sim6\hbar$ and $I= 10\sim20\hbar$.
However, a sharp increase of MOI occurs at $I=6\hbar$ for both of the PSM calculations and the CI-PDFT ones without time-odd interactions.
This disagrees with the experimental observation.
The inclusion of time-odd interactions predicts larger MOI and delays the appearance of the band crossing and, thus, achieves better descriptions of the data.
Note that the enhancements of MOI caused by the time-odd interactions have also been obtained within the framework of cranking and tilted axis cranking relativistic DFTs~\cite{Koenig1993Phys.Rev.Lett.713079--3082,Afanasjev2000Phys.Rev.C031302,Liu2012ScienceChinaPhysicsMechanicsandAstronomy24202424}.

\section{Summary}~\label{sec.V}
In summary, the effects of four-quasiparticle configurations and time-odd interactions are investigated in the framework of configuration interaction projected density functional theory by taking the yrast states of $^{60}$Fe as examples.
The energies and the $B(E2)$ transition probabilities of the yrast states are well reproduced based on the universal PC-PK1 density functional.
Through the analysis of the angular momentum projected states and the probability amplitudes for some important quasiparticle configurations, the dominated roles of the neutron $g_{9/2}$ orbits are emphasised and a weak band crossing caused by the four-quasiparticle configurations at around $I=14\hbar$ is also predicted.
It is found that the inclusion of the time-odd interactions could increase the kinetic moments of inertia and delay the appearance of first band crossing observed in $^{60}$Fe.
This makes the calculated results agree better with the experimental data.

\begin{acknowledgments}
This work was partly supported by the National Key R\&D Program of China (Contracts No. 2017YFE0116700 and No. 2018YFA0404400), the National Natural Science Foundation of China (Grants No. 12070131001, No. 11875075, No. 11935003, No. 11975031, No. 12105004, and No. 12141501), the China Postdoctoral Science Foundation under Grant No. 2020M680183, and the High-performance Computing Platform of Peking University.
\end{acknowledgments}

%

\end{document}